# Mechanism of nucleation and growth near the gas-liquid spinodal

Prabhakar Bhimalapuram[1], Suman Chakrabarty[2] & Biman Bagchi[2,*]

[1]*James Franck Institute, University of Chicago, Chicago, IL 60637, USA*

[2]*Solid State and Structural Chemistry Unit, Indian Institute of Science, Bangalore 560012, INDIA*

* Email: bbagchi@sscu.iisc.ernet.in

## Abstract

**Understanding the mechanism of nucleation of the stable phase inside the metastable parent phase during a first order phase transition has been a subject of outstanding interest in natural science. The problem becomes even more challenging as the spinodal is approached and here the mechanism of nucleation is not clearly understood, with many questions regarding the existence of a free energy barrier and the non-zero value of the surface tension have remained unanswered. In this work, we have undertaken extensive computer simulation studies to probe the molecular mechanism for the onset of instability. We have constructed the multidimensional free energy surface of nucleation as a function of multiple reaction coordinates, both for supercooled Lennard-Jones fluid and for 2- and 3-dimensional Ising models. While the classical Becker-Döring (BD) picture of homogenous nucleation, that assumes the growth of a single nucleus by single particle addition, holds good at low to moderate supersaturation, the formation of the new stable phase becomes more collective and spread over the whole system at large supersaturation. As the spinodal curve is approached from the coexistence**

**line, the free energy, as a function of the size of the largest liquid-like cluster, develops *a minimum at a sub-critical cluster size*. This minimum at intermediate size is responsible for the barrier towards further growth of the nucleus at large supersaturation. As the spinodal is approached closely, this minimum gradually disappears and so does the free energy barrier for the cluster growth. We find the emergence of an alternative free energy pathway (with a barrier less than that in the BD picture) that involves participation of many sub-critical liquid-like clusters and the growth is promoted by coalescence with intermediate sized clusters present in the neighbourhood of the largest cluster. Very close to the spinodal the free energy surface becomes quite flat, *the significance of a critical nucleus is lost* and the classical Becker-Döring picture of nucleation breaks down.**

First order phase transitions usually occur via the formation of a droplet of the stable phase within the metastable bulk phase through an activated process called nucleation.[1-5] The droplet has to grow beyond certain 'critical size' to compensate for the energy required to form the surface. The size and the energy of the critical droplet are often described in terms of the classical Becker-Döring-Zeldovich (BDZ) homogeneous nucleation theory,[1-2] which is based on the assumption that a single droplet of the new stable phase grows by addition of single molecules from the parent metastable bulk phase (at constant chemical potential) to reach the critical size. The growth of the droplet of radius R occurs on the following well-known one dimensional free energy surface

$$\Delta G(R) = -(4\pi/3)R^3|\Delta G_V| + 4\pi R^2\gamma \qquad (1)$$

Where $\Delta G(R)$ is the free energy of formation of a droplet of radius R, $|\Delta G_V|$ is the free energy difference per unit volume between the daughter and the parent phases and $\gamma$ is the surface tension of the interface between them. The above relation gives the following expressions for the size ($R^*$) of the critical nucleus, the free energy barrier ($\Delta G(R^*)$) and the curvature at the top of the barrier

$$R^* = 2\gamma / \Delta G_V$$

$$\Delta G(R^*) = 16\pi\gamma^2 / 3(\Delta G_V)^3 \qquad (2)$$

$$(\delta^2 \Delta G(R)/ \delta R^2)|_{R=R^*} = -8\pi\gamma$$

From the above we get the expected result that both the size $R^*$ and the height $\Delta G(R^*)$ decrease with increase in supersaturation or supercooling and the rate of nucleation increases. While several aspects of this classical nucleation theory have recently been analyzed critically[6,10-15] and different lacunae have been removed, yet the important problem of the mechanism of nucleation at large supercooling, especially that near the gas-liquid spinodal, remains unresolved and somewhat controversial. The well-known droplet model[8] suggests a *kinetic origin* of the spinodal. Mean field theoretic investigations of the spinodal nucleation close to the critical point predicts that the critical size should diverge with increasing quench depth and the process can have 'universal characteristics'.[13,16] A changeover to a *coalescence mechanism* of nucleation near the spinodal was also predicted in a three dimensional Ising model with long range interactions.[14] Recently, Bustillos et al. have constructed the free energy surface for the three-dimensional Ising model as a function of the magnetization and the external field, and found that for states deep in the two-phase region (where the system was expected to be mechanically unstable) the nucleation barrier does become small *but does not disappear,* suggesting that *'there is no spinodal'*![17]

In this work we study spinodal nucleation by first removing the restriction of the existence and growth of a single nucleating cluster and secondly, by calculating free energy as a function of two relevant order parameters for Lennard-Jones fluid. The two reaction coordinates that we use are the total number of particles present in the simulated grand canonical (μVT) ensemble and the 'liquidness' (analogous to the magnetization in the Ising model) of the system. The latter is given by the total number of *liquid-like particles* ($N_{liq}$) identified by its local density. Following the definition by Stillinger[18] and the successful modification by ten Wolde and Frenkel[12] we consider a particle to be liquid-like if it has more than five nearest neighbours within a cut-off distance ($R_c=1.5\sigma$). Liquid-like particles that are connected by neighbourhood (within the cut-off distance) form *liquid-like clusters*. We have used transition matrix Monte Carlo (TMMC) method[19], a recent variant of the widely used umbrella sampling technique, to obtain the three dimensional free energy surfaces at small and large supersaturations.

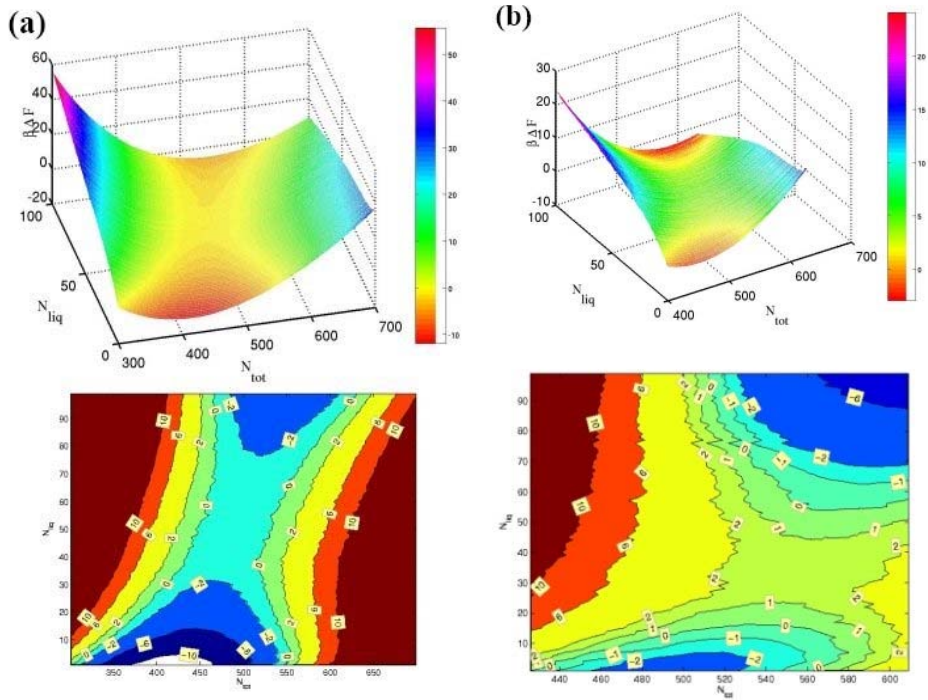

**Figure 1:** The three dimensional free energy surfaces and contour plots computed in grand canonical (μVT) ensemble at reduced temperature $T^*$=0.741 and volume V = (25 σ)$^3$. Activity is defined as ξ=exp(μ/$k_B$T)/λ$^3$, where λ is the de Broglie wave length and μ is the chemical potential. **(a)** (top) 3-dimensional free energy surface and (bottom) 2-dimensional contour diagram for ξ=0.018, which approximately corresponds to supersaturation S~1.8 in the NPT simulation (S=P/$P_c$). **(b)** Same as figure 1(a) at a supersaturation S~2.4 (close to the spinodal) and ξ=0.020.

In **Fig. 1(a)** and **1(b)** we show the calculated free energy surface of formation of liquid-like clusters in a system of Lennard-Jones spheres at two different supersaturations S (given by P/$P_C$, where P is the pressure of the system and $P_C$ is the pressure at coexistence at the same temperature) equal to 1.8 and 2.4. We estimate the spinodal point to be between 2.5-2.6. It is to be noted that this estimate is in close agreement with those of Moody *et al.*[20] (who estimate it to be 2.7 from surface tension calculations) and of Nicolas *et al.*[21] (who estimate it to be 2.2 from equation of state). At intermediate supersaturation (S~1.8), still away from the spinodal, both the activation barrier and the number of liquid-like particles at the barrier are large (**Fig. 1a**). The latter is about 50 (we discuss later that the critical nucleus alone contains nearly all the liquid-like particles) and the activation energy is about 9.5 $k_B$T. On the other hand, near the spinodal (S~2.4), the free energy surface near the saddle is *very flat* (**Fig. 1(b)**). Here the number of liquid-like particles at the saddle is just about 35 and the free energy barrier from the minimum is less than 4 $k_B$T. Importantly, these liquid-like particles are dispersed among several intermediate sized clusters, as discussed below in more detail.

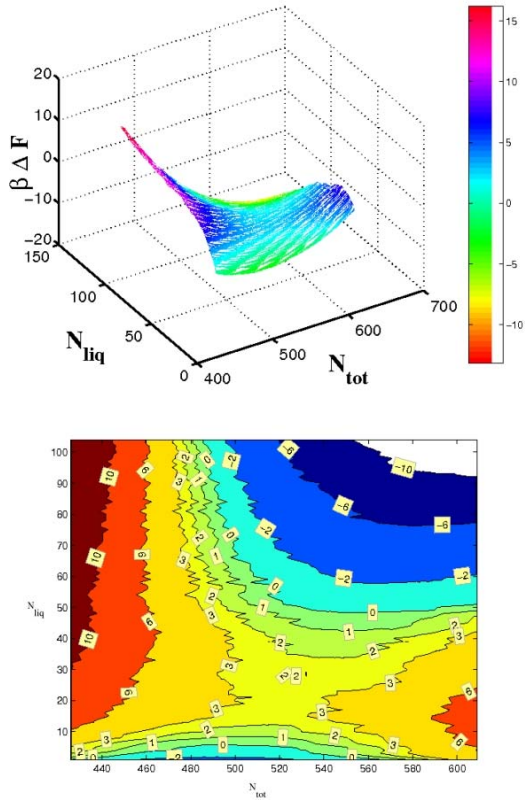

**Figure 1(c)**

To check the basic assumption of the classical nucleation theory, we have computed the free energy surface after restricting the system to have only a single growing liquid-like cluster present in the sea of gas particles. Interestingly, with this constraint *the barrier height becomes higher by more than 1 $k_BT$*. While the critical cluster contains around 25 liquid-like particles, the minimum free energy pathway and the surface near the saddle are quite different in this case (**Fig. 1(c)**). Even in the cluster size distribution of liquid-like particles, computed at two different regions of the free energy surface, one near the saddle region and one in the metastable minimum, we observe that the system at the saddle has a relative abundance of intermediate size clusters as compared to the metastable gas phase. This difference becomes particularly important as spinodal is approached. In contrast, at low supersaturation, the critical cluster contains all of the liquid-like particles and the homogeneous nucleation assumption appears to be valid.

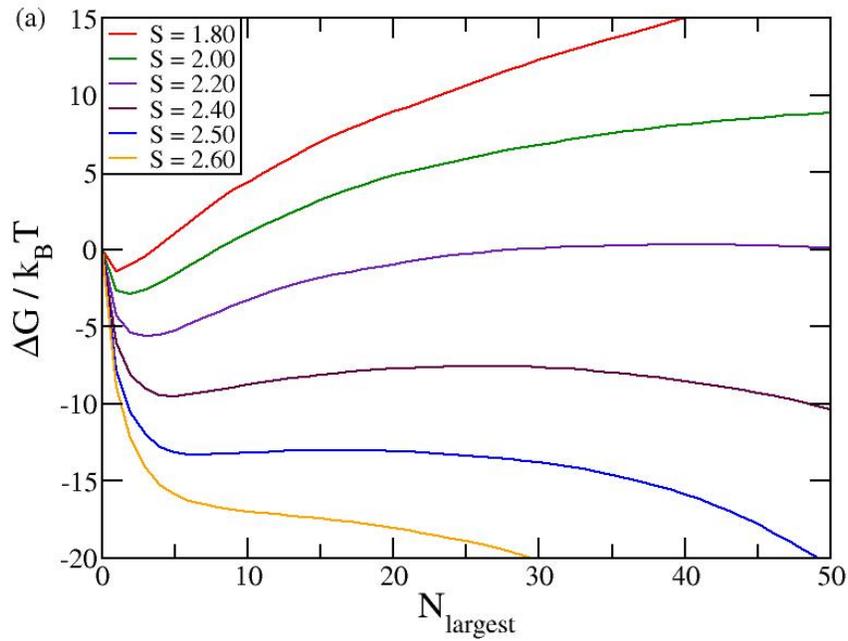

**Figure 2 (a):** Free energy versus size of the largest liquid-like cluster in the NPT ensemble (T*=0.741) for different supersaturation (S=P/P$_c$) across the spinodal in LJ system.

We have carried out extensive simulations in isothermal-isobaric (NPT) ensemble to compute the nucleation barrier for the Lennard-Jones system. Following the usual practice we have chosen the size of the largest liquid-like cluster (N$_{largest}$) in the system to be the appropriate reaction coordinate.[10-12] In **Fig. 2(a)** we show the calculated free energy surfaces for a wide range of supersaturations across the spinodal. We find that at relatively small supersaturation (S=1.53) the classical picture prevails unambiguously. But as the supersaturation is increased, both the size of the critical cluster and the barrier height become progressively smaller and surprisingly *a minimum appears at subcritical cluster size* (N$_{largest}$~5)! At S=2.4 this minimum is very pronounced and essentially responsible for the existence of the nucleation barrier. As the supersaturation is increased towards the spinodal, *the free energy barrier becomes lower* and the minimum at intermediate cluster size becomes deeper and shifts to larger size. Finally it reaches an inflection point at the spinodal beyond which the expected barrier-less continuous growth takes over. To the best of our knowledge, this is the first calculation of the free energy surfaces across the spinodal. The unexpected appearance of the free energy minimum at small liquid-like cluster size indicates the relative abundance and importance of such clusters in the system near the spinodal. Thus, as the degree of supercooling or supersaturation is increased, the growth of the stable phase is not through a single critical nucleus any more and becomes spatially diffuse. The system attains a spinodal character and responds more collectively and *the importance of a critical nucleus is diminished*. This view, as discussed later, is also justified through non-equilibrium quench simulations.

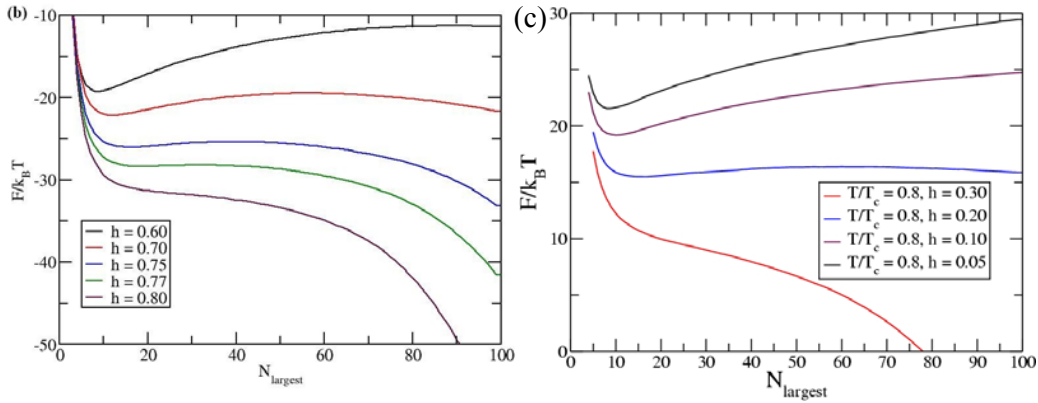

**Figure 2(b):** Free energy versus the size of the largest cluster of up spins for 3-dimensional Ising model at different magnetic fields (h) and T = 0.6 $T_c$. **(c)** Same for 2-dimensional Ising model at different magnetic fields (h) across the spinodal and T=0.8$T_c$.

The above picture remains essentially same in both the 3-dimensional (**Fig. 2(b)**) and the 2-dimensional (**Fig. 2(c)**) Ising models, where the minimum at the intermediate size and the maximum come closer and the free energy surface becomes increasingly flat as the spinodal is approached. This confirms that our observations are rather general. But we must note that the nucleation mechanism can notably differ in these two models since in Ising model the growth proceeds by single spin flips and the clusters cannot physically move as compared to the liquid-like clusters. Based on the present results, we envisage the following picture of spinodal decomposition. The free energy minimum develops at a cluster size where the surface tension contribution starts to become important, in the growth of the cluster, for the first time (surface effects are not important for very small clusters since the density profile is very diffuse). As the supersaturation is increased further towards the spinodal, this minimum shifts to higher cluster size, making the $r^3$ term in **Eq. 1** involving $|\Delta G_V|$, larger than the $r^2$ surface term, leading to the disappearance of the nucleation barrier.

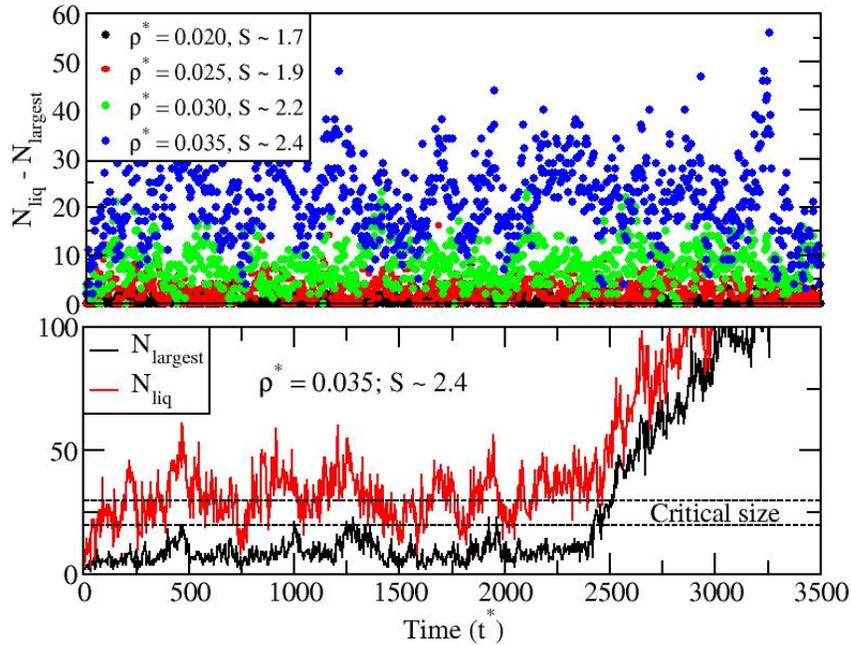

**Figure 3** (top) Evolution of the difference between the total number of liquid-like particles ($N_{liq}$) and the size of the largest liquid-like cluster ($N_{largest}$) subsequent to a temperature quench (from $T^*=1.0$ to $T^*=0.741$). As the same quench is performed in higher densities (nearer to the spinodal curve), the separation between $N_{liq}$ and $N_{largest}$ grows markedly. (bottom) Time evolution of both $N_{liq}$ and $N_{largest}$ is shown for S~2.4. It shows huge fluctuations in both the quantities indicating coalescence of the largest cluster with smaller clusters and break-ups. Note that the largest cluster can break-up even beyond the critical size.

The abundance of the small sized liquid-like clusters suggests the possibility of their coalescence as a mechanism of the formation of the liquid phase. Therefore, we have carried out many non-equilibrium molecular dynamics simulations that consisted of a temperature quench at a constant volume. The volume was chosen such that subsequent to the quench, the system is taken from a stable gas phase to a metastable state near/at/beyond the spinodal curve. In **Fig. 3** we show the difference between the growth of the largest cluster and the total number of liquid-like particles. There are several remarkable features in these growth trajectories near the spinodal. Firstly, *the separation between the largest cluster size and total number of liquid-like particles grows enormously as the supersaturation is increased*. That signifies formation of liquid-like particles throughout the system apart from the largest 'critical nucleus'. Secondly, at S~2.4, the trajectory shows severe fluctuations due to the coalescence and break-ups involving the largest cluster. For even higher supersaturations these jumps are more pronounced and the largest cluster can break-up even beyond the critical size. For vivid picture we include relevant snapshots of the system at different S in **Fig. 4**. This essentially supports the view of the emergence of a spatially diffuse collective mechanism for growth of the stable phase for very high supersaturation.

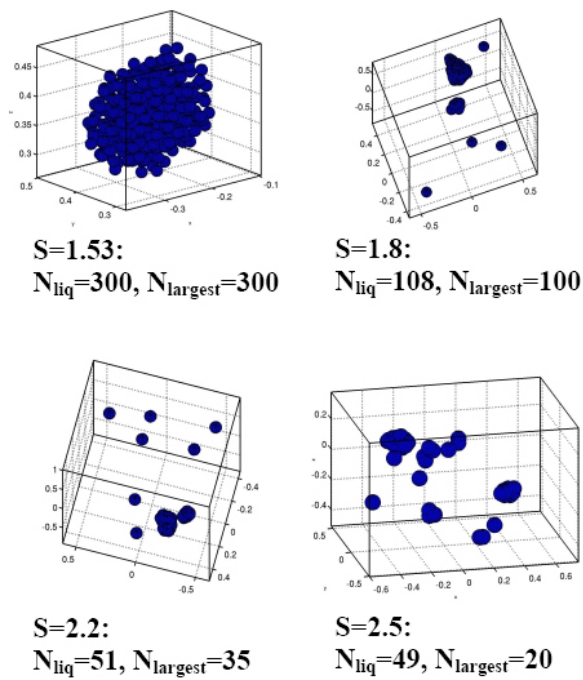

**Figure 4** Snapshots of the system at four different supersaturations (S): They show all the liquid-like particles of the system while the largest cluster is the critical cluster. For higher supersaturations, we find multiple large clusters are forming around the critical cluster and growth of the liquid phase becomes spread over the whole system rather than the single 'critical cluster'.

Our results (including the critical size and the free energy barrier) agree quantitatively with the earlier studies[9,10] at low supersaturation. Previous analyses of the clusters near the critical point of Ising model suggest that the clusters become ramified near the spinodal[20]. In contrast, we find the clusters to be relatively compact for LJ fluid even at S=2.4, with average coordination numbers close to 8 and not ramified. Theoretical studies on spinodal nucleation near the gas-liquid critical point predicted the existence of a diverging length scale. The present study at lower temperatures finds no evidence of any such diverging length. However, we do find the emergence of a large number of liquid-like, sub-critical clusters indicating flatness of the free energy surface which is also a hallmark of all critical phenomena.

Thus it appears from the present simulations that in a macroscopic system, while the critical size may not change much, there can be a macroscopic number of intermediate sized liquid-like particles. We believe that it is the appearance of these intermediate sized liquid-like particles which renders spinodal decomposition a critical phenomena-like character. While the classical nucleation theory remains valid for low to moderate supersaturations, it breaks down at large supersaturation/supercooling. The stable phase starts forming in a spatially diffuse, collective and more continuous fashion. In such a scenario, the question of surface tension does not seem relevant. The disappearance of the free energy barrier for the growth of liquid-like clusters signals the onset of spinodal decomposition.

It is a pleasure to thank to thank Prof. Srikanth Sastry for many helpful discussions and suggestions. This work was supported in parts by grants from DST and CSIR, India. S.C. acknowledges CSIR, India for providing research fellowship.